\documentclass{article}
\fontencoding{T1}
\usepackage{anysize}
\usepackage{cite}
\marginsize{1.8cm}{1.8cm}{1.5cm}{1.5cm}
\usepackage[]{epsfig}
\usepackage{amsfonts}
\usepackage{amsmath}
\usepackage{cite}
\usepackage{pdflscape}

\begin{document}

\title{Dynamic behavior of a Magnetic system driven by an oscillatory external temperature}
\author{M.~Leticia {Rubio Puzzo}$^{1,2,3}$\footnote{Corresponding author. Email address: lrubio@iflysib.unlp.edu.ar}}

\date{\small$^1$Instituto de F\'isica de L\'iquidos y Sistemas Biol\'ogicos (IFLySiB) --- Universidad Nacional de La Plata and CONICET, Calle 59 n.\ 789, B1900BTE La Plata, Argentina \\$^2$CCT CONICET La Plata, Consejo Nacional de Investigaciones Cient\'ificas y T\'ecnicas, Argentina \\$^3$ Departamento de F\'isica, Facultad de Ciencias Exactas, Universidad Nacional de La Plata, Argentina }
\maketitle

\begin{abstract}
 The dynamic effects on a magnetic system exposed to a time-oscillating external temperature are studied using Monte Carlo simulations on the classic 2D Ising Model.
  The time dependence of temperature is defined as $T(t)=T_0 + A \cdot \sin(2\pi t/\tau)$.
Magnetization $M(t)$ and period-averaged magnetization $\langle Q\rangle$ are analyzed to characterize out-of-equilibrium phenomena. 
Hysteresis-like loops in $M(t)$ are observed as a function of $T(t)$.  
The area of the loops is well-defined outside the critical Ising temperature ($T_c$) but takes more time to close it when the system crosses the critical curve. 
Results show a power-law dependence of $\langle Q\rangle$ (the averaged area of loops) on both $L$ and $\tau$, with exponents $\alpha=1.0(1)$ and $\beta=0.70(1)$, respectively.
 Furthermore, the impact of shifting the initial temperature $T_0$ on $\langle Q\rangle$ is analyzed, suggesting the existence of an effective $\tau$-dependent critical temperature $T_c(\tau)$.   A scaling law behavior for $\langle Q\rangle$ is found on the base of this $\tau$-dependent critical temperature.
 
\end{abstract}

\section{Introduction}
\label{sec:intro}

The study of magnetic systems subject to oscillating magnetic fields has attracted the attention of the scientific community in the last decades (see \cite{sides,chakrabarti,keskin,yuksel,liu,ertas} and reference therein). From the point of view of Statistical Physics, phenomena of great interest occur in these systems, such as hysteresis and dynamic phase transitions. The magnetization delay behind the applied field, influenced by relaxation delays, leads to the formation of hysteresis loops, the loop area of which indicates the extent of this delay.

When a ferromagnetic material is exposed to a periodically changing magnetic field over time, it may not respond immediately to the external magnetic influence. This delay arises from a competition between the relaxation time scales of the material's internal dynamics and those of the external magnetic field. 
At high temperatures and for large amplitudes in the periodic magnetic field, the material can gradually synchronize with the external field, although with a noticeable lag. Conversely, this adaptability diminishes under low temperatures and small magnetic field amplitudes.
The occurrence of spontaneous symmetry breaking signals the onset of a dynamic phase transition \cite{chakrabarti}. This is evident in the dynamic order parameter, defined as the temporal average of the magnetization throughout a complete period of the oscillating magnetic field.

As the period of oscillation of the magnetic field decreases, approaching the typical relaxation time of the system, a notable transformation occurs: the hysteresis loop becomes asymmetric around the origin. This asymmetry heralds the emergence of a new thermodynamic phase, a consequence of dynamically broken symmetries due to competing time scales within the system.
This dynamic phase transition is characterized by a dynamic order parameter that exhibits universal power law behavior near the critical point. In particular, the critical exponents reflect those observed in equilibrium transitions, with the temperature and the static external field replaced by the period of the oscillating field and a small bias field, respectively.

In this context, it is worth wondering what effects could occur when such a magnetic system is in contact with a thermal bath whose temperature oscillates over time. In this regard, there are few theoretical studies in this sense \cite{emmert,brandner}, due to the complexity of the phenomenon that will occur completely out of equilibrium.

In this paper, we explore the consequences of subjecting a magnetic system to an oscillating temperature. To achieve this goal we have studied by means of Monte Carlo Simulations, the classic two-dimensional Ising Model when it is subjected to an oscillating temperature.

The work is organized as follows. In Sec. \ref{sec:model}, we give a brief overview of the Ising spin model and Monte Carlo simulations. 
In Sec. \ref{sec:results} we present and discuss the results obtained. Finally, the conclusions are stated in Sec. \ref{sec:concl}.

\section{Model and Simulations}
\label{sec:model}
As it was mentioned in Section \ref{sec:intro} the proposed model is based on the two-dimensional magnetic Ising Model \cite{ising}, whose Hamiltonian (${\cal H}$) 
can be written as
\begin{equation}
{\cal{H}}=-J \cdot \sum _{<i,j>}s_{i}s_{j},
\label{hamis}
\end{equation}
where $s_{i}$ is the Ising spin variable that can
assume two different values $s_i = \pm 1$, the indexes $1\leq
i,j \leq N=L^2$ are used to label the spins, $J>0$ is the coupling constant of the ferromagnet, and the summation is made over nearest-neighbor sites only. 

In the absence of an external magnetic field (as in the present work) and at
low temperature, for $d>1$, the system is in
the ferromagnetic phase and, on average, a majority of spins are pointing
in the same direction. In contrast, at high temperatures the system
maximizes the entropy, thermal fluctuations break the order and
the system is in the paramagnetic phase. This
ferromagnetic-paramagnetic critical transition is a second-order
phase transition and it occurs at a well-defined critical
temperature $T_c$. 
The value of $T_c$ in the 2-d case is obtained from the exact solution proposed by Onsager in 1944 \cite{onsager}, and its value (in terms of $J$) is

\begin{equation}
T_{c}={\frac {2J}{k\ln(1+{\sqrt {2}})}}
\label{eq:Tc}
\end{equation}

The main idea of the present work is the understanding of the effects on a magnetic material when it is introduced on a thermal bath with a time-oscillating temperature. The simplest mathematical approach is to consider
\begin{equation}
T(t)=T_0 + A \cdot \sin(2\pi t/\tau)
\label{eq:tosc}
\end{equation}
where $T_0$ represents the initial temperature, and $A$ and $\tau$ represent the amplitude and period of the oscillation, respectively.

For this purpose, we have simulated --using the standard Monte Carlo method-- an $L\times L$ square lattice of Ising interacting spins (eq. \ref{hamis}) with periodic boundary conditions. The dynamic of the standard model is simulated using well-defined rules, such as the Metropolis rate \cite{metro}, where the probability of flipping a single spin at temperature T is given by
\begin{equation}
W_{standard}=\min[1,\, e^{-\Delta {\cal{H}}/k_B T}],   
\label{eq:wi}
\end{equation},
where $k_B$ is the Boltzmann constant, $T$ is the temperature of
the thermal bath, and $\Delta \cal{H}$  is the difference between the energy of the would-be new configuration and the old configuration.

For the purpose of introducing the time-dependent temperature, we have modified $W_{standard}$ (eq. \ref{eq:wi}) as
\begin{equation}
\label{eq:newrate}
W(t)=\min[1,\, e^{-\Delta {\cal{H}}/k_B T(t)}].
\end{equation}
Certainly, the definition moves the system out of equilibrium, and new phenomena are expected.

The system is left to evolve in time as usual on Monte Carlo simulations, and the time is measured in Monte Carlo time steps (MCs), such as during one MCs all $L^2$ spins of the sample are flipped once, on average.

\section{Results and Discussion}
\label{sec:results}
Monte Carlo simulations were performed on square lattices of size $L^2$, with $64\leq L\leq 1024$, initial temperatures $0.9T_C\leq T_0\leq 1.2T_C$,  the oscillation amplitude was studied in the range $0<A\leq 0.5T_C$, and  $4\leq \tau \leq 256$ (in units of 1/MCs).

The system was initialized at the ordered condition corresponding to $T=0$.
To ensure the system could arrive at a stationary state, we have left evolve the system a time $L^2$ to start the measure of the observables, and we have averaged our results on at least 5000 different realizations of thermal noise. 

The relevant observables analyzed were the absolute magnetization 
\begin{equation}
M(t)=|\sum_{i=1} ^{L^2} s_i|/L^2,
\end{equation}
and the period-averaged magnetization 
\begin{equation}
Q=\frac{1}{\tau} \oint m(t) dt.
\label{eq:q}
\end{equation}

\begin{figure}[!h]
\centering
\epsfig{file=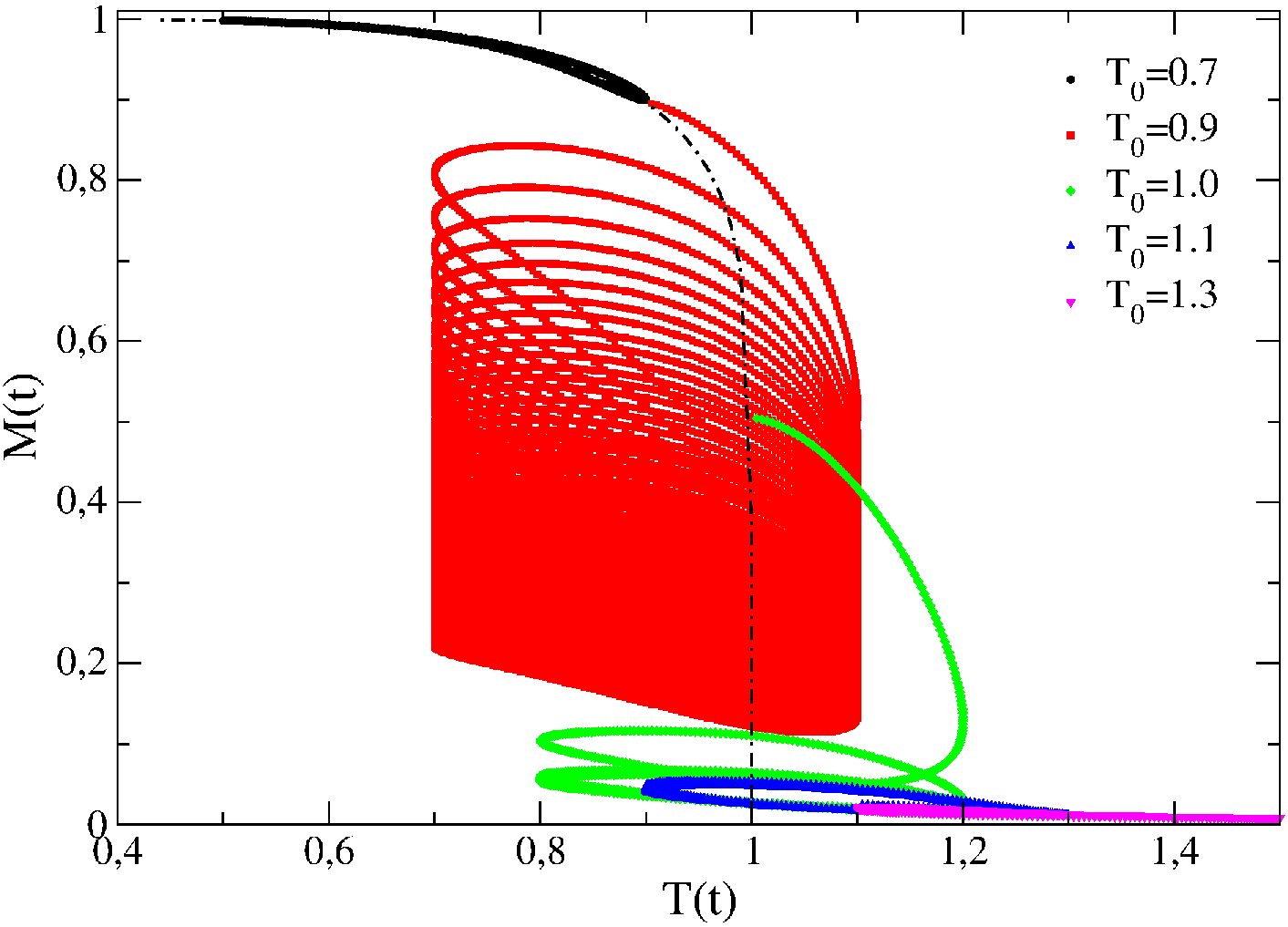,width=10cm,clip=true,angle=0}
\caption{Magnetization vs. Temperature $L=256$, $\tau=64$, and for different initial temperatures $T_0$. The dashed-dotted line indicates the exact solution for the Ising Model phase diagram. }
\label{fig1}
\end{figure}

As a first step, we have analyzed the time behavior of $M(t)$ vs temperature with constant amplitude $A$, and different initial temperatures $T_0$. Figure \ref{fig1} shows the results obtained for $L=256$, $A=0.2$ and $\tau=64$. 
The dashed-dotted line indicates the exact solution for the Ising Model phase diagram, and colored symbols correspond to the data obtained with the time-oscillating temperature. In all cases, a hysteresis-like loop is observed. However, it seems that when the bulk temperature oscillates out of the $T_c$ range ($T_0=0.7T_c$ and $T_0=1.3T_c$) the loop area is close to zero, and it is well defined (the loop closes over itself). On the contrary, when the system crosses the critical curve the loop needs more time to be closed. Similar behavior is observed for different values of $L$ and $\tau$.

\begin{figure}[!h]
\centering
\epsfig{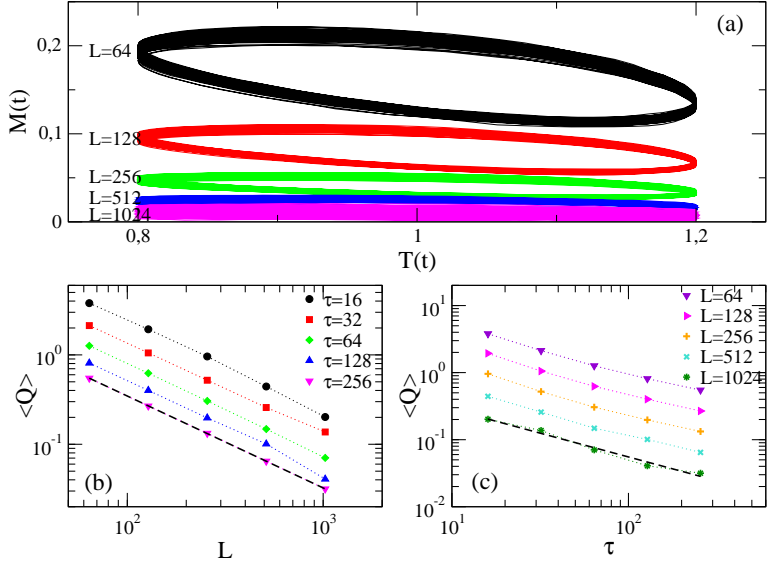}
\caption{(a) Magnetization vs. Temperature for $\tau=64$, $A=0.2$, $T_0=T_c$ and different $L$ as indicated. (b) $\langle Q\rangle$ vs. $\tau$. The dashed line has slope $-1$. (c) $\langle Q\rangle$ vs. $L$. The dashed line has slope $-0.7$ (more details in text).}
\label{fig2}
\end{figure}

From the analysis of the area loops of the magnetization (see as an example Figure  \ref{fig2}(a) for $\tau=64$), it is possible to calculate the value $\langle Q\rangle$ (eq. \ref{eq:q}) averaged over the loops observed and the total number of independent realizations. 
Figure \ref{fig2}(b) shows the obtained results as a function of system size $L$. It is possible to describe the dependence with $L$ as a power-law function of the form $\langle Q\rangle \simeq L^{-\alpha}$, with $\alpha$ an exponent. The best fit of the data gives $\alpha=1.0(1)$ for all the range of $\tau$ studied (dashed line in Figure \ref{fig2}(b)).

Following the procedure described above, we have analyzed the dependence of $\langle Q\rangle$ with $\tau$, for different system sizes $L$. Data obtained is shown in Figure \ref{fig2}(c). As in the previous case, a power-law behavior is observed as $\langle Q\rangle \simeq \tau^{-\beta}$, with $\beta=0.70(1)$ for all the range of $L$ studied (dashed line in Figure \ref{fig2}(c)).

\begin{figure}[!h]
\centering
\epsfig{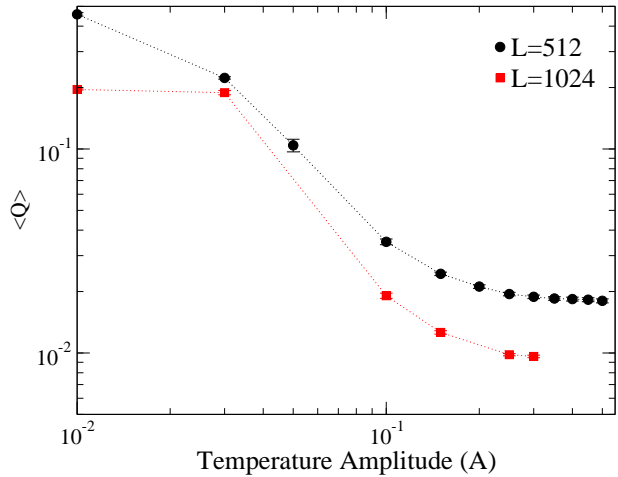}
\caption{Magnetization vs. Temperature Amplitude $A$, for $\tau=256$, $T_0=T_c$ and different $L$ as indicated.}
\label{fig3}
\end{figure}

The effect of varying the amplitude $A$ for fixed $\tau$ and $L$ is shown in Figure \ref{fig3}.
As expected, when the amplitude grows $\langle Q\rangle$ decreases to a constant value $L$-dependent. The plateau observed for larger values of $A$ seems to tend to zero when $L$ increases. This result is compatible with the $L$-dependence of $\langle Q\rangle$ shown in Figure \ref{fig2}(b).

\begin{figure}[!h]
\centering
\epsfig{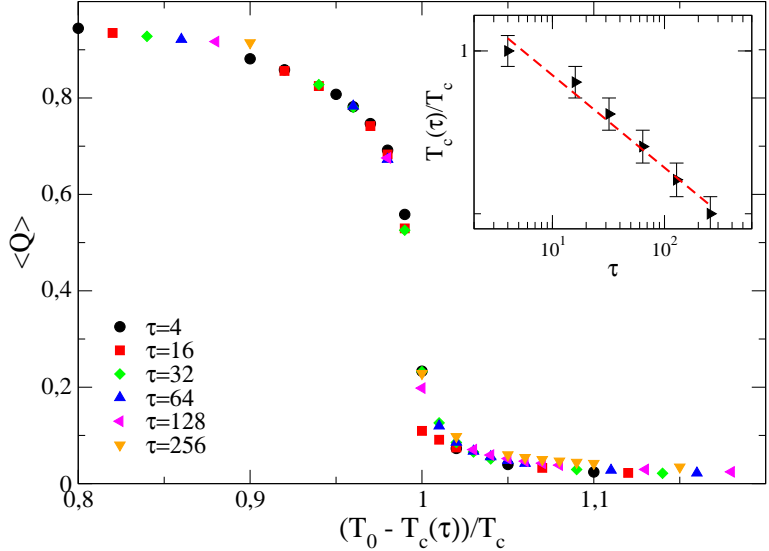}
\caption{$\langle Q\rangle$ vs. the rescaled temperature $\frac{T_0 - T_c(\tau)}{T_c}$ for $L=1024$ and different $\tau$ as indicated. Inset: $T_c(\tau)\approx \tau^{-\gamma}$ vs $\tau$. The dashed line has slope $-0.03$.}
\label{fig4}
\end{figure}

Finally, we have studied the effect on $\langle Q\rangle$ when the initial temperature $T_0$ is shifted. From the data analysis, it is possible to define a $\tau$-dependent critical temperature $T_c(\tau)$ which decreases with the growth of the oscillating period. 
The data obtained suggest that $T_c(\tau)$ have a power-law dependency as $T_c(\tau) \approx \tau^{-\gamma}$, with $\gamma=0.03(1)$ according to the best data fitting (see Inset of Figure \ref{fig4}).
Based on this result it is possible to rescale $\langle Q\rangle$ as a function of the time-oscillating temperature $T(\tau)$ as it is shown in Figure \ref{fig4}. The scaling behavior could be summarized as:
\begin{equation}
\langle Q(\tau) \rangle=f\left(\frac{T_0 - T_c(\tau)}{T_c},L^{-\alpha},\tau^{-\beta} \right)
\label{eq:qtemp}
\end{equation}
where $T_C(\tau) \approx \tau^{-\gamma}$, with $\gamma=0.03(1)$, $\alpha=1.0(1)$, and $\beta=0.70(1)$ .

\section{Conclusions}
\label{sec:concl}

The dynamic effects on a magnetic system embedded into a time-oscillating external temperature are studied.
For this purpose, Monte Carlo simulations on the classic 2d-Ising Model have been performed, where the system temperature has been defined as $T(t) \simeq \sin(2\pi t/\tau)$.
In order to characterize the out-of-equilibrium phenomena, Magnetization $M(t)$ and the period-averaged magnetization $\langle Q\rangle$ have been analyzed.

For $M(t)$ hysteresis-like loops are observed as a function of $T(t)$. The area of the magnetic loops is well defined when the oscillations occur out of the critical Ising temperature value ($T_c$).  On the contrary, when the bath temperature oscillates around $T_c$ the critical curve, the loop needs more time to close on itself. 
This area allows us to study the period-averaged magnetization $\langle Q\rangle$, as a function of system size $L$, and the period $\tau$ and amplitude $A$ of oscillations.
The results obtained indicate that $\langle Q\rangle$ has a power-law dependence with both $L$, and $\tau$, with exponents $\alpha=1.0(1)$, and $\beta=0.70(1)$, respectively.

Finally, the effect on $\langle Q\rangle$ with the initial temperature $T_0$ has been analyzed. The results seem to show that a $\tau$-dependent critical temperature $T_c(\tau)$ could be defined. Based on this, it is possible to construct a scaling law for $\langle Q\rangle$.

In summary, the study reveals interesting dynamic effects on a magnetic system subjected to a time-oscillating external temperature. Hysteresis-like loops, power-law dependencies of $\langle Q\rangle$ on $L$ and $\tau$, and the identification of a $\tau$-dependent critical temperature seem to appear as the consequence of this complex out-of-equilibrium phenomenon.

\section{Acknowledgments}
This work was supported Consejo Nacional de Investigaciones Cient\'ificas y T\'ecnicas (CONICET), Universidad Nacional de La Plata (Argentina), and Agencia Nacional de Promoci\'on de la Investigaci\'on, el Desarrollo Tecnol\'ogico y la Innovaci\'on (Agencia I + D + i) (PICT 2019-2981). Simulations were done on the cluster of Unidad de C\'lculo, IFLYSIB (Argentina).

The author wants to express her deepest gratitude to D.A.M., and to the "Selecci\'on Argentina de F\'utbol" for the third star.

\bibliographystyle{unsrt}
\bibliography{biblio}

\begin{thebibliography}{10}

\bibitem{sides}
S.~W. Sides, P.~A. Rikvold, and M.~A. Novotny.
\newblock Kinetic ising model in an oscillating field: Avrami theory for the
  hysteretic response and finite-size scaling for the dynamic phase transition.
\newblock {\em Phys. Rev. E}, 59:2710--2729, Mar 1999.

\bibitem{chakrabarti}
Bikas~K. Chakrabarti and Muktish Acharyya.
\newblock Dynamic transitions and hysteresis.
\newblock {\em Rev. Mod. Phys.}, 71:847--859, Apr 1999.

\bibitem{keskin}
Mustafa Keskin and Ersin Kantar.
\newblock Dynamic compensation temperatures in a mixed spin-1 and spin-3/2
  ising system under a time-dependent oscillating magnetic field.
\newblock {\em Journal of Magnetism and Magnetic Materials},
  322(18):2789--2796, 2010.

\bibitem{yuksel}
Yusuf Y\"uksel, Erol Vatansever, \"Umit Ak\ifmmode \imath \else \i
  \fi{}nc\ifmmode \imath \else~\i \fi{}, and Hamza Polat.
\newblock Nonequilibrium phase transitions and stationary-state solutions of a
  three-dimensional random-field ising model under a time-dependent periodic
  external field.
\newblock {\em Phys. Rev. E}, 85:051123, May 2012.

\bibitem{liu}
W.I. Liu, Jalal Alsarraf, Amin Shahsavar, Mahfouz Rostamzadeh, Masoud Afrand,
  and Truong~Khang Nguyen.
\newblock Impact of oscillating magnetic field on the thermal-conductivity of
  water-fe3o4 and water-fe3o4/cnt ferro-fluids: Experimental study.
\newblock {\em Journal of Magnetism and Magnetic Materials}, 484:258--265,
  2019.

\bibitem{ertas}
M.~Ertas and M.~Bati.
\newblock Dynamic magnetic properties of spin-7/2 multilayer ising system in an
  oscillating magnetic field.
\newblock {\em Phase Transitions}, 96(3-4):246--257, 2023.

\bibitem{emmert}
Thomas Emmert, Alejandro Cárdenas, and Wolfgang Polifke.
\newblock Low-order analysis of conjugate heat transfer in pulsating flow with
  fluctuating temperature.
\newblock {\em Journal of Physics: Conference Series}, 395(1):012040, nov 2012.

\bibitem{brandner}
Kay Brandner, Keiji Saito, and Udo Seifert.
\newblock Thermodynamics of micro- and nano-systems driven by periodic
  temperature variations.
\newblock {\em Phys. Rev. X}, 5:031019, Aug 2015.

\bibitem{ising}
E.~Ising.
\newblock Beitrag zur theorie des ferromagnetismus.
\newblock {\em Z. Phys.}, 31:253, 1925.

\bibitem{onsager}
L.~Onsager.
\newblock Crystal statistics. i. a two-dimensional model with an order-disorder
  transition.
\newblock {\em Phys. Rev.}, 65(3-4):117, 1944.

\bibitem{metro}
N.~Metropolis, A.~W. Rosenbluth, M.~N. Rosenbluth, A.~H. Teller, and E.~Teller.
\newblock Equation of state calculations by fast computing machines.
\newblock {\em J. Chem. Phys.}, 21(6):1087--1092, 1953.

\end{thebibliography}

\end{document}